# Spinning compact object and chaos in galactic centres

*Ushasee* Paria[1*], *Uditi* Nag[2], *Yeasin* Ali[1,3], and *Suparna* Roychowdhury[1]

[1]Department of Physics, St. Xavier's College (Autonomous), Kolkata-700016, India

[2]School of Physics and Astronomy, Cardiff University, Cardiff CF24 3AA, Wales

[3]Department of Physics, Raja Peary Mohan College, Hooghly-712258, India

**Abstract.** Galactic centres are extremely complex environments, formed primarily of a supermassive black hole (SMBH), which is surrounded by asymmetric mass distributions comprised of nuclear star clusters (NSC), molecular gas, etc forming disks or halos. These elements together exert a gravitational field, which, combined with relativistic corrections from the spin of the SMBH generate strong nonlinear dynamics and chaotic orbits. We model this region using a multipolar expansion potential [1], where the central compact object is denoted by the Artemova–Björnsson–Novikov pseudo-Newtonian potential, which mimics the spin dependence of a Kerr-like BH. The surrounding halo is modelled as an axisymmetric, shell-like mass distribution expanded to the dipole order to induce asymmetry. Previous works have mostly quantified the chaos caused by the BH spin and dipole moment using Poincaré sections, SALI [2] and other such chaos indicators. In this study, we incorporate stability analysis and basins of attraction to provide a global approach to understanding the dynamics of the system. Stability analysis around fixed points gives a deeper context on the local behaviour, while basins of convergence showcase sensitivity to initial conditions and reveal fractal boundaries. Our results show that the BH spin highly reconstructs the equilibrium landscape. Depending on its magnitude and orientation, it can either intensify the chaos induced by the halo asymmetry or simplify the phase space. These discoveries broadly impact our understanding of how relativistic spin effects and multipole moments collectively affect particle motion in galactic centres.

## 1 Introduction

Galactic centres are one of the environments in astrophysics that show the most complex dynamics. Supermassive Black Holes (SMBH) at these centres, like the Milky Way's Sgr A* or the M87* are prime examples of the cause of such complexities. Observational data from JWST, ALMA and VLBI [3], [4], [5] campaigns reveal that these black holes are neither spherically symmetric nor dynamically regular. The SMBHs are encompassed by nuclear star clusters, molecular gas and other mass distributions that are asymmetric in nature. Hence, they participate in shaping a gravitational field that is intrinsically nonlinear. Test objects under the influence of such a field therefore experience chaotic evolutions [6], which often lead to accretion processes, star formation and even jet-launching.

* Corresponding author: ushasee.paria@gmail.com

A major reason behind this complexity is the spin of the central SMBH, which introduces relativistic corrections through frame-dragging and shifts in the location of the innermost stable orbit (ISCO) [7]. Classical Newtonian monopole potentials are not successful in describing such spin-dependent behaviour. To address this problem, several pseudo-Newtonian models have been theorised, among which the Artemova- Björnsson-Novikov (ABN) potential [8], [9] provides a particularly effective approximation by integrating Kerr-like effects, the most important of which is the dependence on the spin parameter a. Moreover, the halo surrounding the SMBH often exhibits vertical asymmetry [2], [10] (due to displaced nuclear clusters), which can be modelled through the addition of a multipolar expansion. Higher order terms are significantly smaller in magnitude, making the dipole term the leading correction to the spherical symmetry in the weak-field region, thus amplifying the nonlinearity of the system [11].

Previous studies have extensively examined how higher-order multipole moments and the spin of a compact object influence the dynamical behaviour of test-particle orbits [10], [11]. These works generally report that the presence of a dipolar moment tends to enhance chaotic dynamics, whereas the spin parameter often exhibits a negative correlation with the degree of chaos. Recent studies usually implement traditional chaos detectors like Poincaré sections, the Smaller Alignment Index (SALI), the Fast Lyapunov Indicator (FLI) [12], the maximum Lyapunov exponent (MLE) [13] and related diagnostics that focus on individual orbits and quantify the chaos of the system.

In this work, we approach the problem through a global lens by analysing the system through its fixed points and final-state indicators, with particular emphasis on Newton–Raphson basins of attraction. This method has been widely used in various branches of physics and nonlinear dynamics, particularly in celestial mechanics and restricted few-body problems [12–14]. However, it has not been extensively employed in systems involving particle dynamics in galactic centres. In particular, the effect of the spin of the central compact object and the surrounding multipolar halo on the behaviour of equilibrium points and on the emergence of fractal basin boundaries remains largely unexplored for such nonlinear gravitational systems.

This approach provides additional insight into the global structure of the phase space and the system's sensitivity to initial conditions, thereby offering a complementary perspective on the onset and nature of chaos in multipolar gravitational fields.

We have categorised our work in this paper in the following way: Section 2 presents the theoretical framework behind the problem, Section 3 explains the methodology behind our approach and provides the results obtained from it and in Section 4 we conclude our results described in Section 3, which is divided into two subsections: in Subsection 3.1 we determine the variation of the number of equilibrium points with the variation of spin, while in Subsection 3.2, we perform Newton-Raphson to find the basins of attraction of the previously obtained equilibrium points.

## 2 Mathematical Formulation

The mass distribution of a galaxy at its core can be described as a multipole expansion, with a central compact object and a galactic bulge around it [6]. For our work, we have considered the expansion up to the dipole term (*D*). The monopole term represents a spherically symmetric Supermassive Black Hole (SMBH), while the dipole term is introduced to account for the vertical asymmetry of the halo density distribution. The spherically symmetric monopole term is substituted by the Artemova-Björnsson-Novikov (ABN) pseudo-Newtonian potential [8] to mimic the Kerr innermost circular orbit (ISCO) effects and study the dynamics of test particles around it. For a=0, the ABN potential depicts the behaviour of a non-rotating (Schwarzschild) black hole. For the Newtonian limit, a=1, the ABN potential becomes the classical -1/R potential. This enables us to probe the full dynamical spectrum—from a non-rotating compact object ((a = 0)) through gradually rotating configurations ((0 < a < 1)) to the Newtonian limit ((a = 1)).)

### 2.1 Description of the potential

The ABN potential in cylindrical coordinates (ρ, ϕ, *z*) is given as [8]:

$$\Phi_{ABN}(\rho, \phi, z) = -\frac{1}{r_1(\beta-1)}\left[\frac{\left(\rho^2+z^2\right)^{\frac{\beta-1}{2}}}{\left(\sqrt{\rho^2+z^2}-r_1\right)^{\beta}} - 1\right] \tag{1}$$

where $r_1$ is the radial position of the event horizon, determined as [9]:

$$r_1 = 1 + (1 - a^2)^{1/2} \tag{2}$$

in units of $GM_{BH}/c^2$. Here $G$ denotes the universal gravitational constant, $M_{BH}$gives the mass of the black hole, $c$ is the speed of light in vacuum and *a* is the rotating parameter (or the spin) [15]. For simplicity, we have assumed $G = c = M_{BH}$=1. Also,

$$\beta = \frac{r_{in}}{r_1} - 1$$

$$r_{in} = 3 + Z_2 - [\left(3 - Z_1\right)\left(3 + Z_1 + 2Z_2\right)]^{\frac{1}{2}}$$

Where
$$Z_1 = 1 + (1 - a^2)^{\frac{1}{3}}[(1 + a)^{\frac{1}{3}} + (1 - a)^{\frac{1}{3}}]$$

$$Z_2 = (3a^2 + Z_1^2)^{\frac{1}{2}}$$

Equation (1) does not possess any ϕ dependence due to the azimuthal symmetry of the system. [2].

The total gravitational potential that acts on a test particle within the halo is given as

$$\Phi_g = \Phi(\rho, \phi, z) + Dz \tag{3}$$

Where $\Phi(\rho, \phi, z)$ denotes the potential arising due to the central compact object and $D$ represents the dipole strength. For our work, we have substituted $\Phi(\rho, \phi, z)$ with $\Phi_{ABN}(\rho, \phi, z)$. To preserve the azimuthal symmetry of the model, we add a term containing the centrifugal force component along the radial direction, given by, $\frac{L^2}{2\rho^2}$ to formulate the total effective potential experienced by the test particle, where L is the conserved angular momentum [2]. Consequently, the net effective potential becomes

$$U_{eff} = \Phi_g + \frac{L^2}{2\rho^2}$$

Implying
$$U_{eff} = \Phi(\rho, \phi, z) + Dz + \frac{L^2}{2\rho^2} \tag{4}$$

### 2.2 Newtonian Dynamics

So, the equations of motion become

$$\dot{\rho} = p_\rho \tag{5a}$$

$$\dot{p}_\rho = -\frac{\partial U_{eff}}{\partial \rho} \tag{5b}$$

$$\dot{z} = p_z \tag{5c}$$

$$\dot{p}_z = -\frac{\partial U_{eff}}{\partial z} \tag{5d}$$

Where we have adapted the convention $GM_{BH} = 1$ and $c$=1.

## 3 Results and Discussion

In this section, we first determine the equilibrium points for this system and analyse their stability. Then we construct Newton-Raphson basins of attraction of the fixed points to determine the nature of convergence of initial conditions towards them with the variation of spin. The detailed discussion on the methods and the presentation of our acquired results follows in the succeeding subsections:

### 3.1 Stability Analysis

Equilibrium points, or fixed points are the locations in the (ρ, z) plane where the effective potential is stationary, satisfying the condition $\nabla U_{eff} = 0$. The fixed points of the system are found by solving the equilibrium equations [10]

$$\frac{\partial U_{eff}}{\partial \rho} = \frac{\partial U_{eff}}{\partial z} = 0 \tag{6}$$

By solving equation (6), we locate the equilibrium points using the Newton-Raphson solver, in accordance with standard literature on non-linear dynamics and numerical analysis [16], [17]. We determine the stability of each equilibrium point by analysing the eigenvalues of the Hessian matrix of $U_{eff}$, consistent with the idea in [18]. For a particular equilibrium point, if both the eigenvalues are greater than zero, it is taken as stable. If at least one eigenvalue is less than zero, it becomes an unstable fixed point. This follows from the classical theory of conservative two-dimensional potentials [6].

Thereupon, we proceed to plot the variation of the total number of fixed points obtained for every spin with respect to the spin parameter $a$ for 200 values between 0 and 1.

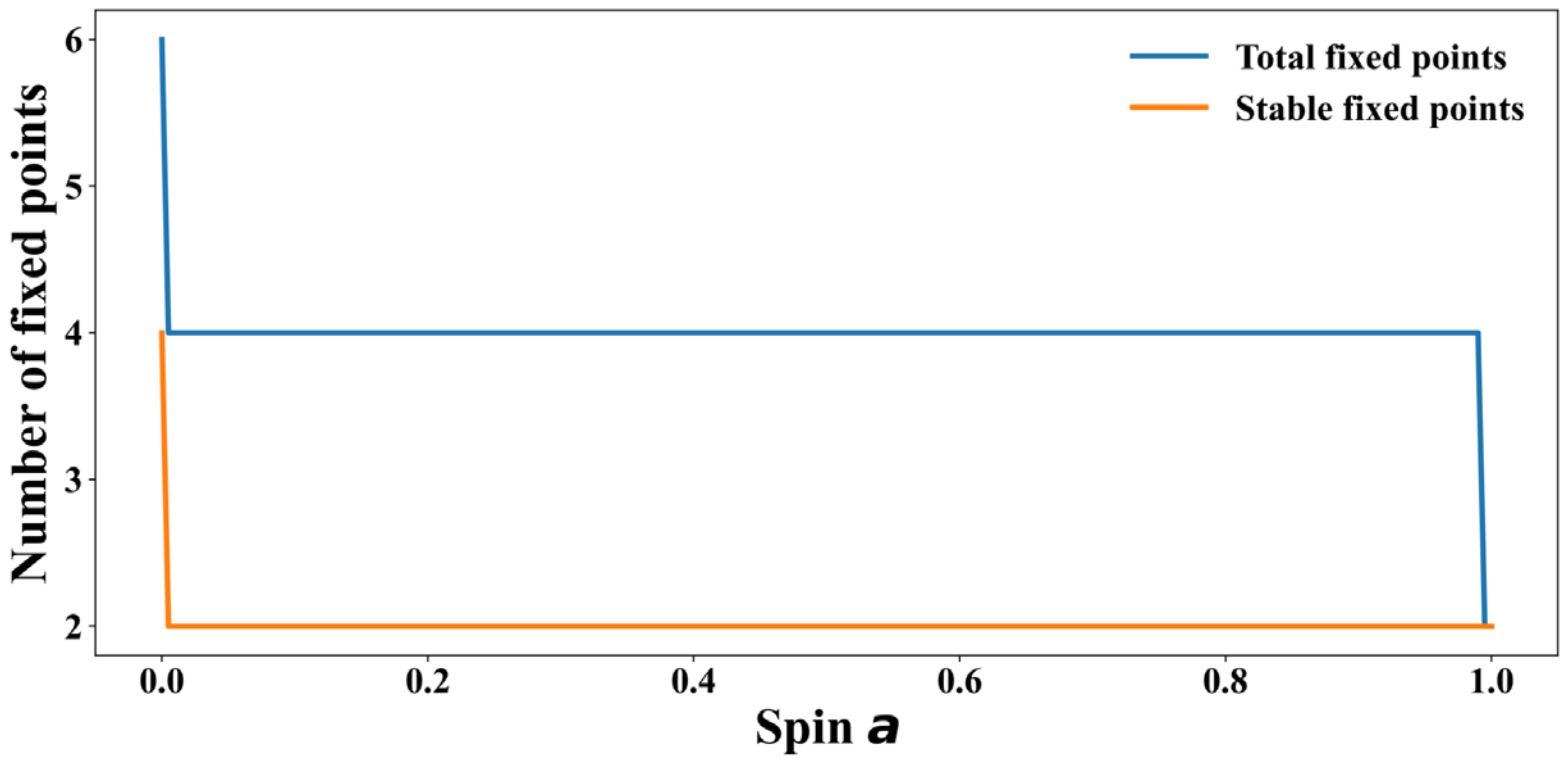

**Figure 1.** We constructed the plot for the variation of the number of fixed points with change in spin parameter $a$ from 0 to 1 for 200 values. The blue line represents the total number of fixed points, while the orange line shows the number of stable fixed points.

In Figure 1, we have shown the evolution of the number of fixed points with the variation of the spin parameter $a$. The figure suggests that the general trend is a reduction of fixed points with an increase in the spin parameter. For $a = 0$, the total number of fixed points is 6, out of which 4 are stable. This number drastically reduces to 4 (with the number of stable fixed points becoming 2) right after $a$ becomes greater than zero. The number of fixed points, both total and stable, remains constant for a long time, right till $a$ reaches 1, where only 2 stable fixed points remain. We see, however, that the introduction of spin to the non-rotating compact object only causes an initial reduction in the number of equilibrium points, but the magnitude of the spin has no effect on it until the system reaches the Newtonian limit.

Taking into account the behaviour of the number of fixed points with the increase of spin, we take a detailed view of it by characterising the nature of the potential for some specific spin values. To achieve this, we plot the two contours of the effective potential $U_{eff}$, *i.e.*, we plot the equations (6) along with the fixed points for those particular spin values, and mark their stability [18].

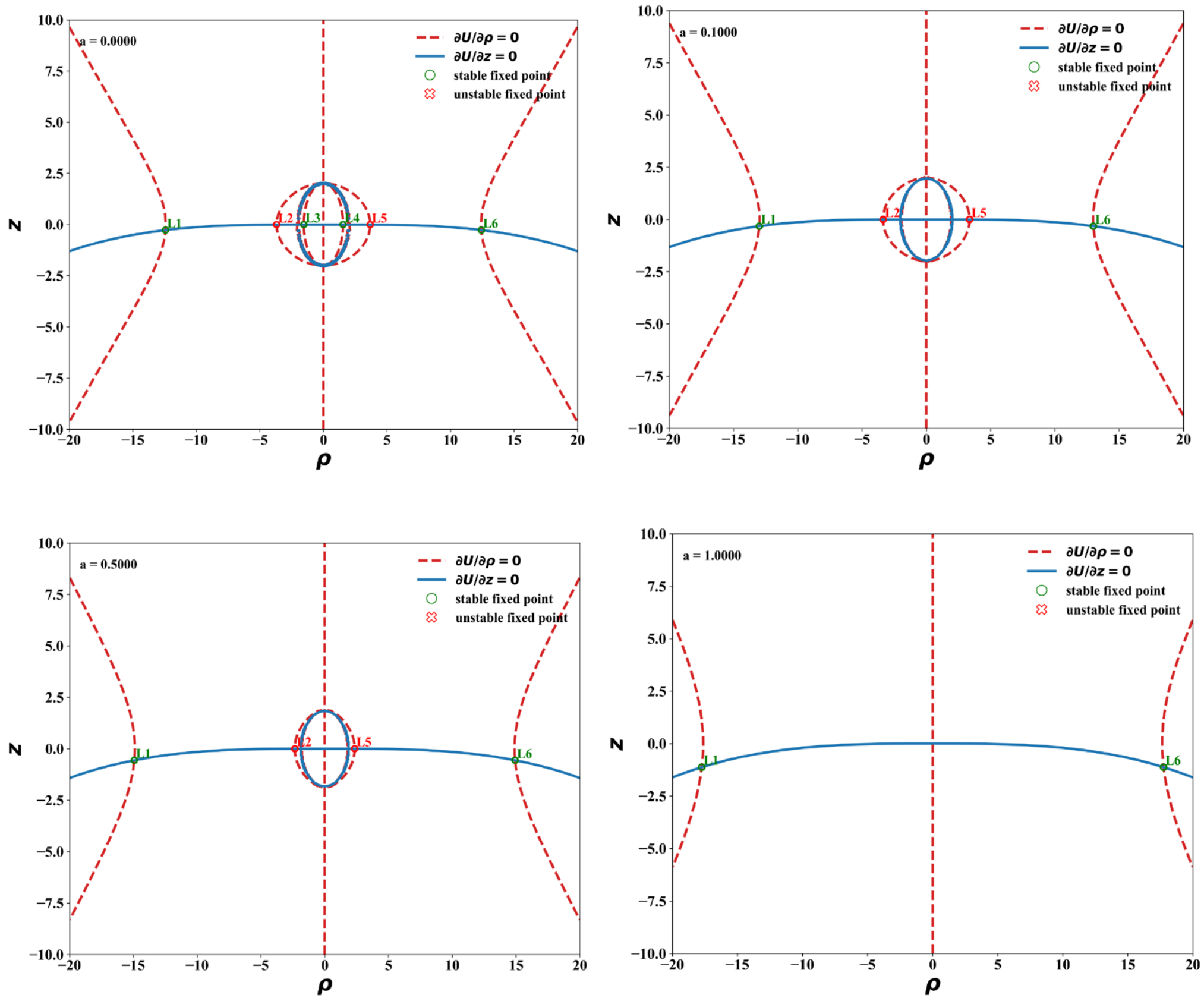

**Figure 2.** We plotted the positions of equilibrium points along with the contours in the *ρ*-*z* plane for spin values *a*=0.0, *a*=0.1, *a*=0.5 and *a*=1.0, respectively. Red dotted lines resemble the ρ- derivative of the effective potential, while the blue solid lines indicate the z-derivative. Green dots resemble stable fixed points, red dots are for unstable ones.

Figure 2 shows the contours of the effective potential $U_{eff}$ plotted in the ρ-z plane for four specific values of the spin parameter, along with the respective fixed points. The intersections of the contours resemble the position of the fixed points. This validates the results obtained from Figure 1: we indeed start with 6 fixed points (L1, L2, L3, L4, L5, L6) at $a = 0$ , with L1, L3, L4 and L6 being stable. Then for $a = 0.1$, we see that two of the previously stable fixed points, L3 and L4 disappear, leaving two stable fixed points L1 and L6 and two unstable fixed points L2 and L5 respectively. This behaviour persists for $a < 1$, as confirmed by the plot for $a = 0.5$. Finally, at $a = 1$, there are only two stable fixed points L1 and L6 remaining. It is also noteworthy that with the increase of spin, L1 and L6 tend to migrate outwards, while L3 and L4 eventually move inwards symmetrically as spin increases. Thus, in the limit $0 < a < 1$ , the spin parameter $a$ does not influence the number of fixed points, but it tends to shift their positions.

## 3.2 Newton-Raphson Basins of Convergence

We once again employ the Newton-Raphson method to numerically solve the equations (6). We begin with an initial condition $(\rho_0, z_0)$ on the configuration plane, then we carry on this iterative procedure till we reach any attractor (fixed point) of the system, with some preassigned accuracy. If the iteration ends up on one of the fixed points, then we conclude that the method has converged for that specific initial condition. However, all initial conditions might not necessarily converge to any attractor in the system [13]. Those initial conditions which converge to a particular final fixed point together form the basins of convergence [14], also known as the Newton-Raphson basins of attraction, or attracting regions/domains.

To obtain the basins of convergence, we performed the following: Firstly, we specified a fine grid of 1024 × 1024 initial conditions uniformly distributed over the coordinate (ρ, z) plane and conducted a Newton-Raphson iteration. The iteration was stopped when it arrived at an accuracy of $10^{-9}$, and we grouped all the initial conditions in terms of (ρ, z) that lead to one particular equilibrium point. For our work, the maximum number of iterations $N_{max}$ was set to 200, which proved sufficient to capture the true dynamical structure, avoiding over-smoothing of basin boundaries [13]. We now try to determine how the spin parameter $a$ influences the Newton-Raphson basins of attraction.

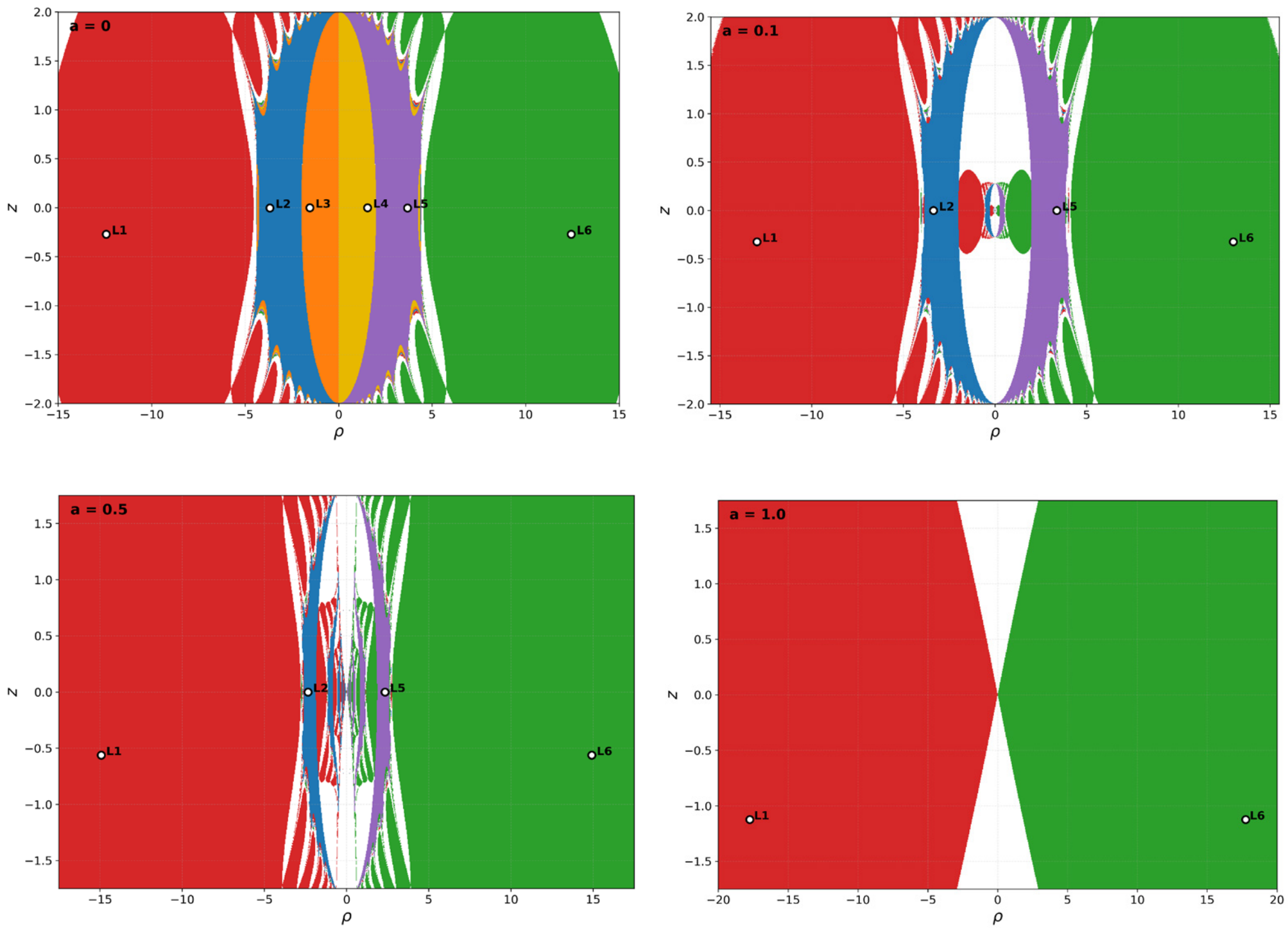

**Figure 3.** We plotted the basins of convergence in the (*ρ*-*z*) plane for the four cases, *a*=0.0, *a*=0.1, *a*=0.5 and *a*=1.0, respectively. The white circles denote the fixed-point centres, and each fixed point is assigned a specific colour. The colour code denotes the basins of the attractors L1 (red), L2 (blue), L3 (yellow), L4 (orange), L5 (violet), L6 (green) and non-converging points (white).

In Figure 3, we display the Newton-Raphson basins of convergence for four values of the spin parameter $a$ of the central compact object. For $a = 0$ (the Schwarzschild case), we observe all six equilibrium points (L1-L6) with highly intricate fractal boundaries separating their basins of attraction. The basin structure exhibits characteristic features of chaotic dynamics: colored finger-like intrusions penetrating deeply into neighbouring basins, particularly visible at the top and bottom boundaries, where red (L1) and green (L6) fingers intrude into the interior basins. This means that the choice of a starting point $(\rho_0, z_0)$ of the Newton-Raphson method inside these fractal domains is extremely sensitive. Even if we make a small change in choosing the initial conditions, it results in a totally different final attractor, and therefore, it becomes very difficult to predict the convergence beforehand. [12]. For $a = 0.1$ the central fixed points L3 and L4 disappear completely, leaving only four attractors (L1, L2, L5, L6). There is a large central white zone where the initial conditions fail to converge to any attractor. L3 and L4 either merge with other fixed points or disappear completely. L2 and L5 possess small basins near the equatorial plane, while L1 and L6 are dominant. Red and green island-like regions appear in the central zone and the whole basin structure is replicated continuously as we move towards the origin. The fractal finger-like intrusions at the top and bottom are still present with further modifications. This suggests that even small values of spin fundamentally restructure the equilibrium landscape. For $a = 0.5$ the structure with four attractors persists, but the central white region expands significantly. The basins of L2 and L5 become even narrower and more elongated, appearing as thin colored strips in the central region. The fractal finger-like structures at the boundaries become thinner and more widely spread, with extensive white space between them. Red and green intrusions penetrate deeply towards the centre but are interrupted by large non-converging regions. This indicates that at this intermediate spin value, the system is undergoing significant dynamical changes.. For $a = 1$ only two attractors (L1, L6) remain and the basin structure becomes much simpler with smooth and clearly distinguishable basin boundaries, indicating the predictability of dynamics. The white space is confined to a very narrow region along the central boundary. The plot shows high simplification as the system approaches the Newtonian limit with only two competing potential wells.

Correlating the four cases, we observe that the basins of L1 and L6 steadily grow with increasing spin, eventually occupying almost the entire configuration space at $a = 1$. Their boundaries transition from highly fractal and complex at low spins through a transitional regime with substantial white space at intermediate spins to smooth and regular at the Newtonian limit. The persistence of fractal structures in the outer regions, even at $a = 0.1$ indicates that some chaotic dynamics survive after the loss of L2 and L5, even though the central region is already simplified.

## 4 Conclusion

This paper aimed to characterise the influence of the spin parameter $a$ of the central compact object on the motion of the test particle under the effect of the pseudo-Newtonian gravitational potential of that compact object and the asymmetric mass distribution around it. We graphically obtained the variation of the number of equilibrium points with the change in $a$ and visualised the contours of the potential along with the location of the fixed points for certain spins. The results showed that for a non-rotating compact object, the number of fixed points in our chosen (ρ, z) plane came out to be 6, which reduced to 4 when the compact object was characterised by the Artemova- Björnsson-Novikov potential, and ended up being 2 in the Newtonian limit. We also performed a stability analysis and showed that the number of stable equilibrium points reduced to a constant value of 2 for $a > 0$, initially being 4 at $a = 0$. We also visualised a shift in the outer stable fixed points farther away from the origin, and in the inner unstable fixed points towards the origin, *i.e.*, the position of the central compact object. Hence, the introduction of the spin reduced the number of equilibrium points and also caused their migration as it increased in value.

We also constructed the basins of convergence of the 6 equilibrium points by implementing the Newton-Raphson method. For the case $a = 0$, the configuration (ρ, z) plane is a complex combination of convergence basins and fractal boundaries, which is a signature of chaotic dynamics. This proves high sensitivity in those regions towards slight changes in initial conditions, leading to unpredictable final states (attractors). The fractal structures, manifested as colored finger-like intrusions of one basin deeply penetrating the neighbouring basins indicate regions of extreme sensitivity to initial conditions. Even a small introduction of spin ($a = 0.1$) reduces the number of fixed points by two. This significant change proves that adding even a very small amount of spin to the compact object fundamentally reorganises the phase space. The central region where L3 and L4 previously existed becomes dominated by white space, indicating that these fixed points have either merged with others or disappeared. This sudden transition between $a = 0$ and $a = 0.1$ is one of the most striking features of our analysis and suggests that spin-induced effects on particle dynamics are highly nonlinear. The extremal limit eliminates L2 and L5, leaving only two dominant attractors L1 and L6. The basin structures are extensively simplified as the fractal boundaries vanish entirely, replaced by a smooth, almost linear boundary between the two basins. We also note that the fixed points and Newton-Raphson basins are symmetric in nature, owing to the symmetry of the effective potential. An accuracy of $10^{-9}$ was implemented in constructing the basins. We noted that for higher accuracies, many points failed to converge and the NR basins did not give any conclusive results. However, the number and position of fixed points remained the same. So, this adopted tolerance can be considered optimal for constructing suitable basins of convergence as well as maintaining the numerical accuracy of the computation. In addition, the maximum number of iterations was varied from 500 to 1000, but no significant difference in the percentage of convergence or shape of basins was found. We wish to see if decreasing the accuracy by a few more orders and increasing the maximum number of iterations tenfold would give better results.

Taking into account the outcomes of our studies, we come to the general conclusion that the spin parameter has a significant role in the manifestation of the chaotic nature of the system. The introduction of spin reduces the number of attractors in the system, and it also shapes the basins of attraction and affects their sizes. These findings have important conclusions for particle dynamics around compact objects: 1. Even slowly rotating black holes show fundamentally different phase space structure compared to non-rotating cases, as shown by the fewer number of equilibrium points and simplified basin structures. 2. Sensitive dependence on initial conditions at $a = 0$ suggests that particle trajectories may be inherently chaotic near equilibrium regions. 3. The extensive white space at intermediate spins indicates that certain rotations may result in highly complex dynamics, as many particle trajectories remain undecided for extensive periods. 4. Newtonian limit shows the simplest, most predictable dynamics.

Future prospects of this work would include: (1) performing a fine-resolution scan of spin values in the range $0 \leq a \leq 0.1$ to precisely identify the critical spin value where L3, L4 disappear; (2) extending the stability and basin of attraction analysis for higher-order multipoles of the asymmetric mass distribution (quadrupolar and octopolar terms) to characterize a more realistic galactic model; (3) computing fractal dimensions of basin boundaries as a function of spin to characterize the transition from chaotic ($a = 0$) to regular ($a = 1$) dynamics quantitatively. The non-linear nature of the ABN potential and its derivatives, particularly close to the Kerr-like object, presents significant numerical challenges for the Newton-Raphson iteration in the central region, manifesting as extensive white space where convergence fails. Further investigation is required to determine whether these non-converging regions represent genuinely chaotic trajectories influenced by the strong relativistic corrections present in the ABN potential or they are just numerical limitations of the iterative method when applied to the highly nonlinear derivatives of the effective potential. There is an abrupt transition between $a = 0$ and $a = 0.1$ where two central fixed points vanish. This reflects that we need to deeply investigate this small range of spin to determine the exact critical spin value where the reduction occurs and to identify if a bifurcation occurs. If so, we also need to classify the bifurcation type (saddle-node, transcritical, or pitchfork). Additionally, analysing the intermediate spin regime ($0.3 \leq a \leq 0.7$), where white space is the most prominent, could reveal whether the system passes through a transitional state from maximum dynamical complexity to a simplified structure as we move towards the Newtonian limit. Such studies would help us understand how relativistic spin effects captured by pseudo-Newtonian potentials and multipolar mass distributions jointly govern the dynamical behaviour around galactic centres and also predict the accretion processes around supermassive black holes.